\documentstyle[aps,twocolumn,epsf,floats]{revtex}
\begin{document}
\wideabs{
\title{Negative Resistance and Local Charge-Density-Wave Dynamics}
\author{H. S. J. van der Zant and E. Slot}
\address{Department of Applied Sciences and DIMES, Delft University of Technology, Lorentzweg 1, 2628 CJ Delft, The Netherlands}
\author{S.V. Zaitsev-Zotov and S.N. Artemenko}
\address{Institute of Radioengineering and Electronics,
Russian Academy of Sciencies, Mokhovaya 11, 103907 Moscow, Russia.}
\date{Submitted to PRL}
\maketitle

\begin{abstract}
Charge-density-wave dynamics is studied on a submicron length scale
in NbSe$_3$ and o-TaS$_3$. Regions of negative absolute
resistance are observed in the CDW sliding regime at sufficiently low
temperatures. The origin of the negative resistance is attributed to
the different forces that the deformed CDW and
quasi-particles feel: the force on the CDW is
merely caused by a difference of the {\em electric} potentials, while
the quasi-particle current is governed by a difference of the {\em
electrochemical} potentials.

\end{abstract}

\draft
\pacs{PACS numbers: 71.45.Lr, 71.45.-d, 72.15.Nj}
}

A periodic modulation of the conduction electron density is commonly observed in low-dimensional conductors \cite{Peierls}. This charge-density-wave (CDW) state is the ground state in various inorganic and organic materials with a chain-like structure, giving rise to remarkable electrical properties \cite{Gruner,Thorne}.
A particularly interesting feature of the CDW is its collective
transport mode, somewhat similar to superconductivity
\cite{Frohlich}. Under an
applied electric field, CDWs slide along the crystal, giving rise to a
strongly nonlinear conductivity. Since even a small amount of disorder pins
the CDWs, sliding occurs only when the applied electric field exceeds a
certain threshold.

In metallic and superconducting devices, reduction of sizes has
revealed a variety of new mesoscopic phenomena.
For CDW conductors, the mesoscopic regime has only been studied for small transverse dimensions~\cite{BZZN2,GillSize,ThorneSize} because samples of (sub)micron sizes in the chain direction could not be fabricated in
a controlled way. Consequently, many aspects of microscopic CDW dynamics are still unknown.
Nevertheless, some first studies on a micron-scale revealed interesting mesoscopic features
related to the CDW phase distribution~\cite{ZZPokGill,PokZZ}.
More recently, there have been attempts to fabricate artificial submicron CDW devices.  Latyshev {\em
et al.}~\cite{latyshev00} made antidot arrays and found indications
that the current-conversion mechanism changes at the submicron
scale. A similar conclusion has been drawn by Mantel et
al.~\cite{mantel00}. In submicron CDW wires, they observed an unexplained
size-effect for the phase-slip voltage if probe spacings are in the order of a few micron.

\begin{figure}
\vskip -1.9cm
\epsfysize=14cm
\centerline{\epsffile{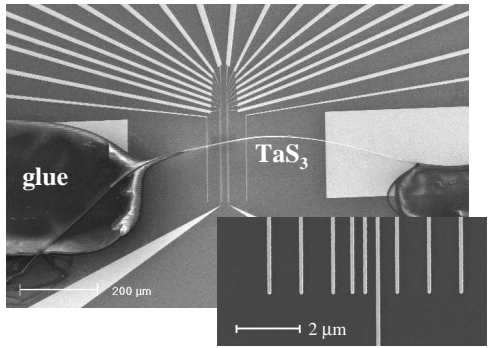}}
\vskip -7.1cm
\caption{
A thin TaS$_3$ crystal on top of an array of voltage probes to study CDW dynamics on submicron length scales.
The spacing between the big (current) pads on either side of the picture is 0.5~mm.
The inset shows an enlargement of the main figure with 9 voltage probes that are 100~nm wide; the smallest distance between adjacent probes is 300~nm. Each sample has two of these probe-sets that are separated 12~$\mu$m from each other.
}
\vskip -.3cm
\label{foto}
\end{figure}

In this paper, we report on CDW dynamics on a (sub)micron scale.  We
present current-voltage characteristics ($IV$'s) recorded on
high-quality NbSe$_3$ and TaS$_3$ crystals with probe spacings in the
(sub)micron range (see the inset of Fig.~\ref{foto}).
On these short length scales $IV$ curves appear
to vary strongly from segment to segment.  For some segments the
absolute resistance may even become negative.

Experiments were carried out on single NbSe$_{3}$ and o-TaS$_3$ crystals with
cross sections of 0.2 to 1~$\mu$m$^2$.  Both materials have a very
anisotropic, chain-like structure \cite{Gruner}.  NbSe$_3$ exhibits CDW
transitions at $T_{\rm P1}=145$~K and
$T_{\rm P2}=59$~K. 
At low temperatures a small portion of the conduction electrons remains
uncondensed, providing a metallic single-particle channel.
In contrast, in o-TaS$_3$ all electrons condense into
the CDW state. As a
result, the linear resistance shows semiconducting behavior below the transition
temperature of 220~K.

A common technique to contact small CDW whiskers consists of putting the crystals on top of metal probes that are evaporated on an insulating substrate.  Then a droplet of glue (ethyl cellulose dissolved in ethyl acetate) is put on the CDW crystals to keep them fixed on the metal probes.
In previous studies the smallest probe widths were on the order of 2~$\mu$m and
their smallest separations were typically 10~$\mu$m.  By using standard e-beam
lithographic techniques, we have fabricated an array of gold wires that are
50~nm high and 100~nm wide.  The smallest probe separation is 300~nm as
illustrated in the inset of Fig.~\ref{foto}.
It is important to note that to study
microscopic CDW dynamics, our results show that both the probe width and separation must be sufficiently small.

Electrical contact between o-TaS$_3$ and the gold wires has only been obtained
after heating the crystals to 120-130~$^{\rm o}$C for several minutes to one
hour~\cite{BZZN1}. During this annealing step, sulfur that has
accumulated at the surface oxidizes leaving behind a clean interface. For NbSe$_3$, we heat the samples so
that the thin crystals do not start floating in the glue solvent.  When the
substrate is heated to 80$^{\rm o}$C the solvent evaporates quickly, giving the
crystals no opportunity to float.

A series of measurements has been performed to characterize the
crystals.  Cross sections ($S$) have been determined from
measuring the resistance $R$ for segments with different
separation $L$ at room temperature. Here, $L$ is defined as the
distance between the middle of two voltage probes. We find that
$R$ scales perfectly with $L$. Cross sections are then calculated
using the literature values of the room-temperature resistivity:
$\rho = 2$~$\Omega \mu$m for NbSe$_3$ and $\rho = 3$~$\Omega \mu$m
for o-TaS$_3$.  Another test involves the measurement of Shapiro
steps when biasing the samples with both a dc and ac drive.  At
120~K, we obtain complete mode locking for both crystals
indicating their high quality.  As expected, we find that the
step-width scales linearly with frequency yielding sample cross
sections that compare very well with the values obtained from the
resistance measurements (within 7~\%).  Threshold fields ($E_T$'s)
of the crystals also point at a good sample quality. At 120~K,
$E_T=0.58$~V/cm for a NbSe$_3$ crystal of $S=0.2~\mu$m$^2$ and
$E_T=1.23$~V/cm for a TaS$_3$ crystal of $S=0.5~\mu$m$^2$.

We have systematically studied $IV$ characteristics of all
possible segments. We first discuss the results obtained on
o-TaS$_3$ crystals and show $IV$'s on micron-sized segments
obtained in the normal four-probe configuration: Current is
injected at the large gold pads (see Fig.~\ref{foto}) and various 
probe
pairs measure the voltage between them. Measurements on o-TaS$_3$
are performed in the temperature range of 94-220~K.

\begin{figure}
\vskip -2.5cm
\epsfysize=13cm
\centerline{\epsffile{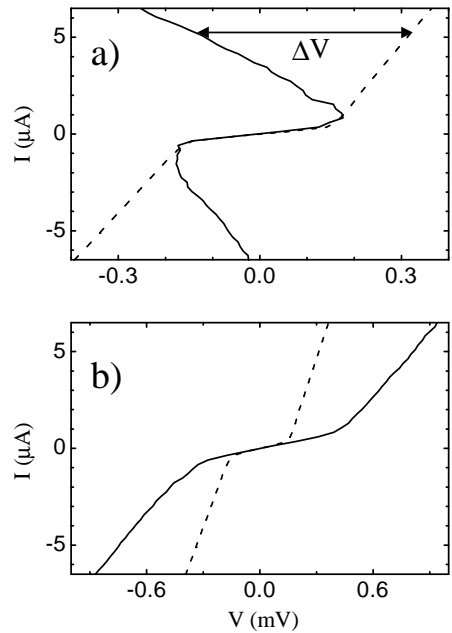}}
\vskip -2cm
\caption{
Two current-voltage characteristics of a TaS$_3$ crystal with a cross section of 0.5~$\mu$m$^2$. The curves are taken at 120~K on adjacent 1~$\mu$m-long segments. Dashed lines correspond to a measurement on a 31.6~$\mu$m long segment for which the voltage has been scaled by a factor of 1/31.6. They represent the expected, averaged nonlinear CDW behavior. In a) the absolute value of the resistance becomes negative for high positive bias. The deviation $\Delta V$ from the expected behavior is linear in $I_{CDW}$.  The curve in b) shows less CDW current at a given field, i.e., it is less nonlinear. When adding up the two curves one approximately recovers the expected, average CDW behavior.
}
\label{IV}
\end{figure}

When $L$ is larger than about 10~$\mu$m, we always observe the expected behavior for CDWs as indicated by the dashed lines in Fig.~\ref{IV}a and b. On a micron scale,
however, the shape of the $IV$ varies from segment to segment.
Some small segments show the same nonlinear behavior as observed in the large
segments.  Other segments may yield an $IV$ that shows a negative differential
resistance (NDR). Occasionally the absolute resistance may even become negative (see the
positive current branch in Fig.~\ref{IV}a for $I>3.5$~$\mu$A),
indicating that the
moving CDW pumps charge carriers in the direction opposite to the applied field.
A small adjacent segment exhibits an $IV$ curve that is less nonlinear and it
appears to have a somewhat higher threshold (Fig.~\ref{IV}b). However, these
differences average out when measuring on larger length scales and deviations
between various $I(E=V/L)$-curves are small if $L>10$~$\mu$m.

Taking a closer look at the curves in Fig.~\ref{IV}a, one sees that the deviation
($\Delta V$) from the expected $IV$ (dashed line) is proportional to $I_{CDW}$. We can express this deviation with a parameter $\alpha$ defined as $\alpha=\Delta V /(R_0 I_{CDW}$), where $R_0$ is the linear resistance around $V=0$.  From
Fig.~2a, we find that $\alpha \approx 0.36$ for the positive current branch and
$\alpha \approx 0.2$ for the negative current branch.

We have also recorded $IV$'s at different temperatures below the
Peierls temperature of 220~K. At 200~K, we find no unusual behavior:  all $IV$
characteristics can approximately be collapsed on one $I(E)$-curve.  For the
segment of which the 120~K data are shown in Fig.~\ref{IV}a, NDR sets in at 160 K with $\alpha =0.17$ and 0.11 for the positive and negative current branch respectively. As temperature decreases, N(D)R becomes more pronounced.
For example, at 94~K, $\alpha \approx 0.96$ for the positive current branch and $\alpha \approx 0.6$ for the
negative current branch.  Situated 20~$\mu$m from this segment, another segment
with $L=0.8$~$\mu$m also starts to develop N(D)R at temperatures below 100~K.
Again the adjacent segments show less nonlinear behavior with a somewhat
higher threshold field.

In total, we have studied four different o-TaS$_3$ samples.  All of them showed
NDR in one or two segments, each time involving different probe pairs.
A trend seems to be that the thinner the sample the
more pronounced N(D)R shows up.
We have found no indications of serious damage of the crystals (cracks) in the NR region when measuring the electrical properties.
In all four samples, $R_0$ and $E_T$ scale with $L$ within error margins for all segments.
We also measured the two-probe resistances between all probe pairs at several temperatures.
Again no unusual features were found.
The two-probe resistance is generally two orders of magnitude larger than the four-probe resistance.
They only vary by a factor five from each other and no trend was observed for the probes that show NDR (e.g.  no systematically lower contact resistance).

N(D)R has also been detected in two NbSe$_3$ samples when measuring $IV$ curves on the micron scale.
In one sample, the absolute resistance became negative.
The conditions to find N(D)R in NbSe$_3$ appear to be more stringent.  We only observe it below the second Peierls transition ($T_{\rm P2}$) and in the close vicinity ($\sim 1~\mu$m) of a current contact.

Summarizing our experimental results, we conclude that negative resistance (NR) has been observed in both o-TaS$_{3}$ and NbSe$_{3}$.  In all cases it is a local effect occurring on a micron scale.  In NbSe$_{3}$, however, NR has only been observed close to a current contact, whereas in o-TaS$_{3}$ it has been observed when current contacts are far away. Also, NR has only been observed in NbSe$_{3}$ for $T<T_{\rm P2}$ and in o-TaS$_{3}$ for $T<160$~K.  The effect is more pronounced as the temperature is lowered indicating that the quasi-particle density (resistivity) plays a role in the observation of NR.
Lastly, all $IV$'s are asymmetric with respect to interchanging the positive and negative current branches.

In early reports on CDW dynamics on macroscopic distances, NDR has also been observed in NbSe$_3$ crystals~\cite{hall84} and in TaS$_3$~\cite{ZZPokGill,BZZN1}.
In contrast, for $L>5$~$\mu$m, we have never seen NDR in our four-terminal measurements.
NDR has also been observed on large distances in partly irradiated CDW crystals by Latyshev {\it et al}.~\cite{latyshev89}.  Their qualitative
explanation involves strong CDW deformations in the strain profile
near the boundary of the irradiated and non-irradiated part.

We are not aware of any measurement that shows NR
in CDW samples and we believe that it is not related to the
observations on NDR reported by other groups.  The main reason for
this is the local character of the NR in our samples.  In the
remainder of this paper, we provide a qualitative explanation for the
observation of NR in CDWs.

The basic ingredient of our model is that the CDWs and the quasi-particles 
are driven by different forces. At low temperatures, when the
quasi-particle density is much smaller then the density of the
condensed electrons, the force exerted on the CDW is mostly
related to a difference of the {\em electric} potentials
$\Phi$~\cite{AVK}. In contrast, the quasi-particle current is
governed by a difference of the {\em electrochemical} potential,
i.e., by the voltage drop $V=U(x_1)-U(x_2)$.

\begin{figure}
\vskip -1cm
\epsfysize=10cm
\centerline{\epsffile{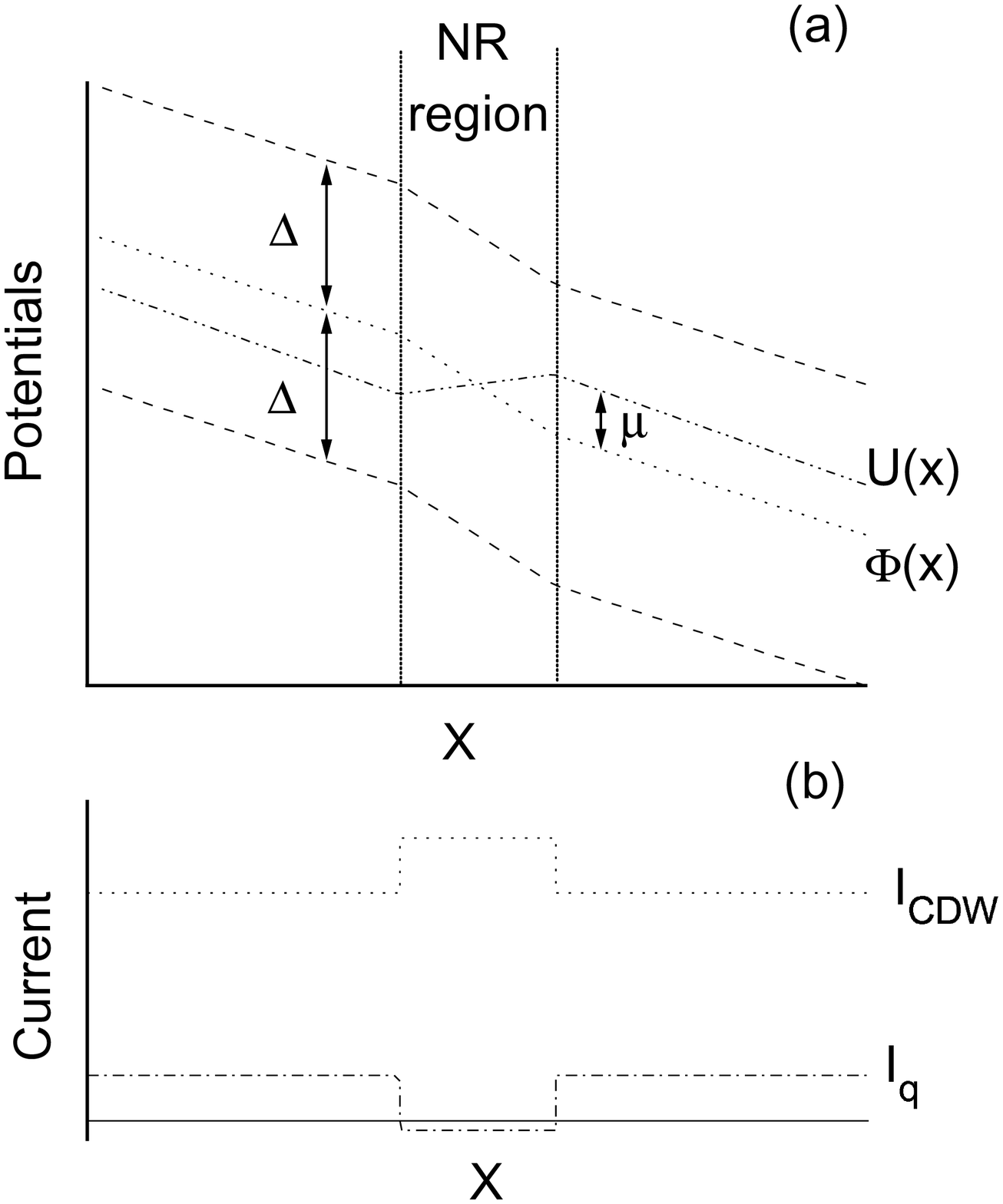}}
\vskip -1.1cm
\caption{(a) Band bending and (b) current distributions
around the NR region.  $\Phi$ coincides with the middle of the Peierls gap under the chosen calibration of the electrostatic potential.
Edges of the Peierls gap are represented by the dashed lines. }
\vskip -.4cm
\label{fig:explanation}
\end{figure}

Starting from the experimental data, we can draw the potential and
current distributions along the sample defining the conditions for
which NR can be observed. First of all, the slope, $dU/dx$, in the
NR region is opposite to the mean slope of $U(x)$ along the sample
(the solid line in Fig.~\ref{fig:explanation}a). The
quasi-particle current $I_q$ is equal to $\sigma_q dU/dx$ with
$\sigma_q$ is the linear conduction per unit length. It therefore
also has a sign opposite to that in the rest of the sample
(Fig.~\ref{fig:explanation}b). Since the total current,
$I=I_q+I_{CDW}$, is the same along the sample, the CDW current
$I_{CDW}$ in the NR region must be larger than in the rest of the
sample (Fig.~\ref{fig:explanation}b). At the same time as
illustrated in Fig.~\ref{fig:explanation}a, one sees that the force
on the CDW is larger in the NR region because the gradient of the
{\em electric} potential $E=-d\Phi/dx$ is larger here. Finally,
one should keep in mind that the difference ($U-\Phi$) defines
shifts of the chemical potential ($\mu$) with respect to the
midgap position.  These shifts are related to CDW phase
deformations, i.e., $\mu = (U-\Phi) \propto
d\varphi/dx$~\cite{BZZN2,AV,IPN}, where $\varphi$ is the CDW
phase.

We now concentrate on the physical conditions that lead to the
potential and current distributions shown in
Fig.~\ref{fig:explanation}.  If we assume that $K$ in the relation
$I_{CDW}=K\dot\varphi$ is the same along the sample, then the jump
in $I_{CDW}$ (see Fig.~\ref{fig:explanation}b) corresponds to a
larger CDW phase velocity $\dot\varphi$ in the NR region.  Such a
variation of $\dot\varphi$ must be provided by phase slip, but the
corresponding CDW deformations in Fig.~\ref{fig:explanation}a have
the wrong sign~\cite{zz89}. This contradiction means that NR
results from a variation of $K\equiv I_{CDW}/\dot\varphi$ rather
than from a variation of the CDW velocity only.

As $\dot\varphi$ has to be the same along the sample, we assume
that $I_{CDW}=(1+\alpha) K\dot\varphi $ in the NR region (the
origin of $\alpha$ will be discussed later), and that
$I_{CDW}=K\dot\varphi$ for the rest of the sample. Since the total
current along the sample is constant, one has
$K\dot\varphi+I_q=(1+\alpha)K\dot\varphi+V/R_0$. The voltage drop
in NR region is then given by:
\begin{equation}
V=R_0\left(I_q-\alpha I_{CDW}\right),
\label{eq:Vnr}
\end{equation}
where $R_0=l/\sigma_q$ with $l$ the length of the NR region.
It is obvious from Eq.~\ref{eq:Vnr} that for $I_{CDW}>I_q/\alpha$ the
sign of $V$ is negative. Note that $\Delta V=\alpha R_0 I_{CDW}$ is
the deviation from the regular $IV$ curve introduced earlier.

There is, however, a limitation for the CDW to keep the same phase velocity
$\dot\varphi$ through an NR region. Phase slippage is expected to set in as soon as the CDW deformation reaches its
limiting value: $\mu_{\max}=V_{ps}/2e$, where $V_{ps}$ is
the phase-slip voltage~\cite{AVK}. Therefore,
$\Delta V$, is limited by the phase-slip voltage and $\Delta
V_{\max}=(I_{CDW}/I)V_{ps}$. Thus, NR cannot be observed in segments
having $l>l_{\max}\sim \alpha V_{ps}/E_T$, and for total currents
$I>V_{ps}\sigma_q/l\alpha$. For typical values ($V_{ps}\sim 1$~mV,
$\alpha\sim 0.1$ and $E_T\sim 1$~V/cm) one gets $l_{\max}
\sim 1$~$\mu$m, in good agreement with the experimental data.

A variation of $K = (2e/\pi) N_c (1-b)$~\cite{AV89} and thus a nonzero $\alpha$ can be related to a change in the number of the conducting chains in the sample's cross section ($N_c$), or to a change in the parameter $b$. Here, $b$ describes the reduction of
``condensed'' electrons with increasing temperature and contains the so-called ``back-flow current'' which depends on the scattering times.
A plausible explanation might therefore be that increased scattering in a NR region -due to for example a macroscopic defect, e.g. a line dislocation- leads to a quasi-particle current that flows in a direction opposite to that in the rest of the sample. Hall measurements on K$_{0.30}$MoO$_3$ have shown~\cite{forro86} that back flow currents produce typical $\alpha$-values 
of $0.1-0.2$, in good agreement with our high-temperature data.

Theoretical estimates based on a microscopic
approach~\cite{AV89,Art97} indicate that the contribution of $b$ is
too small to account for the observed low-temperature $\alpha$-values of $\sim 1$.  However, at low temperatures the
experimentally obtained values are likely to be
overestimated.
The measured values of $\alpha$ are very sensitive to a correct
determination of $R_0$.  Furthermore,
in TaS$_3$ at $T<100$~K, $R_0$ greatly depends on the CDW deformations
which undoubtedly occur near defects. In that case $R_0$ is smaller than
expected from the simple Arrhenius law, $R_0\sim \exp(\Delta/T)$,
where $\Delta$ is the Peierls gap~\cite{Tako}.

More detailed calculations on the local CDW dynamics are needed to explain all experimental details.
These should also address the asymmetry of the $IV$ curves in the NR region. It is most likely caused by the absence of symmetry between electron-like and hole-like excitations and the built-in CDW deformations that shift the chemical potential.

We thank Robert Thorne for providing the NbSe$_3$ and o-TaS$_3$
crystals.  We further acknowledge discussions with Nina
Markovi\'{c}, Yuri Latyshev and Yuli Nazarov. This work was
supported by the Netherlands Organization for Scientific Research
(NWO), by the Netherlands Foundation for Fundamental Research on
Matter (FOM), by the Russian Foundation for Basic Research
(project 01-02-17771) and the Russian program on physics of
nanostructures (project 97-1052). HSJvdZ was supported by the
Dutch Royal Academy of Arts and Sciences
(KNAW).

\end{document}